# Do Mendeley readership counts help to filter highly cited WoS publications better than average citation impact of journals (JCS)?[1]

Zohreh Zahedi[1], Rodrigo Costas[2] and Paul Wouters[3]

[1a&b]z.zahedi.2@cwts.leidenuniv.nl; [2]rcostas@cwts.leidenuniv.nl; [3]p.f.wouters@cwts.leidenuniv.nl

[1a,2,3] *CWTS, Leiden University, P.O. Box 905, Leiden, 2300 AX (The Netherlands)*

[1b] *Department of Knowledge & Information Sciences (KIS), Faculty of Humanities, Persian Gulf University, Bushehr, 7516913817 (Iran)*

**Abstract**
In this study, the 'academic status' of users of scientific publications in Mendeley is explored in order to analyse the usage pattern of Mendeley users in terms of subject fields, citation and readership impact. The main focus of this study is on studying the filtering capacity of Mendeley readership counts compared to journal citation scores in detecting highly cited WoS publications. Main finding suggests a faster reception of Mendeley readerships as compared to citations across 5 major field of science. The higher correlations of scientific users with citations indicate the similarity between reading and citation behaviour among these users. It is confirmed that Mendeley readership counts filter highly cited publications (PPtop 10%) better than journal citation scores in all subject fields and by most of user types. This result reinforces the potential role that Mendeley readerships could play for informing scientific and alternative impacts.

**Conference Topic**
Altmetrics

**Introduction**
Mendeley is a popular reference management tool and a rich source of readership metrics for scholarly outputs, used by more than 2.5 million users[2]. This platform collects a wide variety of different metadata[3] for each publication saved by the different types of users in their individual library. Among these metadata, statistics about 'academic status', 'discipline' and 'country' provide useful information on the typologies of users of scientific publications in Mendeley.

Mendeley has different coverage and presence across different fields of science (Zahedi, Costas & Wouters, 2014). A moderate correlation between Mendeley readership and citation counts has been observed for different sets of publications from different fields showing that Mendeley readership counts reflect similar but (perhaps) also other types of impact (Thelwall et al., 2013; Haustein et al., 2013; Zahedi, Costas & Wouters, 2014; Mohammadi & Thelwall, 2014). Also, a weak correlation among number of authors, departments, institutions and countries and readership and citation counts for WoS publications has been observed (Sud & Thelwall, in press; Thelwall & Maflahi, in press). Research on users showed that the majority of Mendeley users per publication are PhDs and students. However, one important limitation with Mendeley data on the analysis of users was the data restriction caused by the reporting of only the three most common user types per publication. Full data on users are necessary in order to properly determine the readership patterns among types of users (Zahedi, Costas & Wouters, 2013 & 2014; Haustein & Larivière, 2014; Mohammadi et al., 2014).

---

1. This paper presented at the 15th International Conference on Scientometrics and Informetrics (ISSI), 29 Jun- 4 July, 2015, Bogazici University, Istanbul (Turkey)
2. http://blog.mendeley.com/start-up-life/mendeley-has-2-5-million-users/
3. See: http://apidocs.mendeley.com/home/user-specific-methods/user-library-document-details

The new Mendeley API provides data on all typologies of readers per publication. This means that 100% of all the users per publication are now fully reported[4]. This study represents one of the first approaches to the analysis of Mendeley readerships based on statistics per publication from all users. We overcome the main limitation of previous studies which were limited to restricted Mendeley users statistics.

In this paper, the usage patterns of the different Mendeley users based on their 'academic status'[5] by fields, citation and readership impact are studied. Also, we analyse the extent to which Mendeley readerships correlate with the number of citations and across 5 major fields of science in the Leiden Ranking (LR). An important focus of this study is on studying the filtering capacity of Mendeley readerships compared to journal citation scores in detecting highly cited publications. Therefore, particular attention will be paid to the extent to which highly cited outputs can be distinguished by these different impact indicators. Similarly, potential differences among Mendeley users in detecting highly cited publications will be also explored. The concrete objectives and research questions of the paper are the following:

O1: To study the general distribution of Mendeley readerships over WoS publications
    Q1. What is the distribution of Mendeley readerships across LR fields and by different users?
O2: To study the relationship of Mendeley readerships with bibliometric indicators
    Q2. Are there any differences in correlation by different Mendeley users and across LR fields?
O3: To investigate the ability to identify highly cited publications by Mendeley readerships in contrast to journal citation impact indicators
    Q4. Which one of these impact indicators can better filter the WoS highly cited publications across LR fields and by different users?

**Data and Methodology**
For this study, we used a dataset of 1,196,421 Web of Science (WoS) publications from the year 2011 with Digital Object Identifiers (DOI). DOIs were used as the basis to extract readership metrics through the Mendeley REST API in mid-October 2014. The data from Mendeley has been matched with the CWTS in house WoS to add citation data. Citations have been calculated up to 2014.

Although Mendeley has released the full statistics for all the typologies of the users per publications through its API, some Mendeley user statistics are still missing from some publications[6]. These publications were excluded from the analysis due to their unclear reader counts and types. Limiting the dataset to articles and reviews, a final set of 977,067 publications received 12,418,426 total readerships[7] and 6,882,632 total citations. Comparing the ratios of mean citation score per publication (MCS) and mean readerships per publication (MRS), we also find higher MRS (12.7) than MCS (7.04). The actual number of the different types of Mendeley users per publication has been calculated as well as several bibliometrics indicators. Precision-recall analysis (Waltman & Costas, 2014) has also been performed, considering 5 major fields of science as represented in the Leiden Ranking (LR)[8].

---

[4].according to William Gunn in the 1:Am altmetrics conference in London (September 2014) www.altmetricsconference.com/
[5.] These are the different types of users in Mendeley (i.e. PhD students, Professors, Post doc, researchers, Students (under graduates and post graduates), Librarians, Lecturers, Other Professionals and Academic and non-Academic researchers) who have saved publications in their individual libraries. This information allows us to identify users of scientific publications but this information is not free of limitations. For example, it is not clear whether the academic status of the users is updated regularly or how to distinguish users who could belong to more than one category (e.g. a librarian who is also a PhD student).
6 . There are 144,8496 publications with missing readership statistics. These publications have been saved in Mendeley but since their readership counts are missing, they are excluded from the analysis.

**Analysis and Results**

*General distribution of Mendeley readerships by major fields of Science and by Mendeley users*

Table 1 shows that Biomedical & health sciences (37%) have the highest share of publications with readerships while Mathematics and computer science (8%) have the lowest share. In terms of readership density (i.e. MRS scores) the Life & earth sciences have the highest values (17.5) followed by the Social science & humanities (17), Biomedical & health sciences (14.4) and Natural sciences & engineering (9.7). Mathematics and computer science (9.4) exhibit the lowest readerships density. Also, on average, all fields show higher MRS scores than MCS scores. This could be explained by the relative early publication year (2011) of publications, which could still need some time to get their optimum levels of citations, while in terms of social media, the uptake is normally faster (Haustein et al, 2013), although we still lack information on the obsolescence and time patters of readerships for publications.

**Table 1. Mendeley readerships distribution across 5 major fields of science in LR**

| LR Main fields of all Publications | P | % | TCS | % | MCS | TRS | % | MRS |
|---|---|---|---|---|---|---|---|---|
| Biomedical & health sciences | 419,693 | 37 | 3617563 | 44 | 8,6 | 6051206 | 39 | 14,4 |
| Natural sciences & engineering | 322,009 | 28 | 2362700 | 29 | 7,3 | 3119704 | 20 | 9,7 |
| Life & earth sciences | 204,392 | 18 | 1469979 | 18 | 7,2 | 3572266 | 23 | 17,5 |
| Social sciences & humanities | 105,827 | 9 | 422046 | 5 | 4,0 | 1795194 | 12 | 17,0 |
| Mathematics & computer science | 90,813 | 8 | 332946 | 4 | 3,7 | 857319 | 6 | 9,4 |
| **Total** | | 100 | | 100 | | | 100 | |

Total Citation Score (TCS); Total Readership Score (TRS); Mean Citation Score (MCS); Mean Readership Score (MRS)

Figure 1 shows the proportion of readerships by the different types of Mendeley users across the LR fields. Although there are some differences across the fields, in general we find that PhD and students are the most common types of users while Lecturers and Librarian are the least common types of users across all LR fields.

---

[7]. We have found some inconsistencies in the counts of readerships. There is a difference between the sum of total readership counts reported by Mendeley (i.e. as they come directly from the readership count provided by Mendeley) and the sum of the individual Mendeley readerships by the different users (calculated by ourselves). (12,418,426 - 12,412,305=6121 differences)

8. http://www.leidenranking.com/ranking/2013

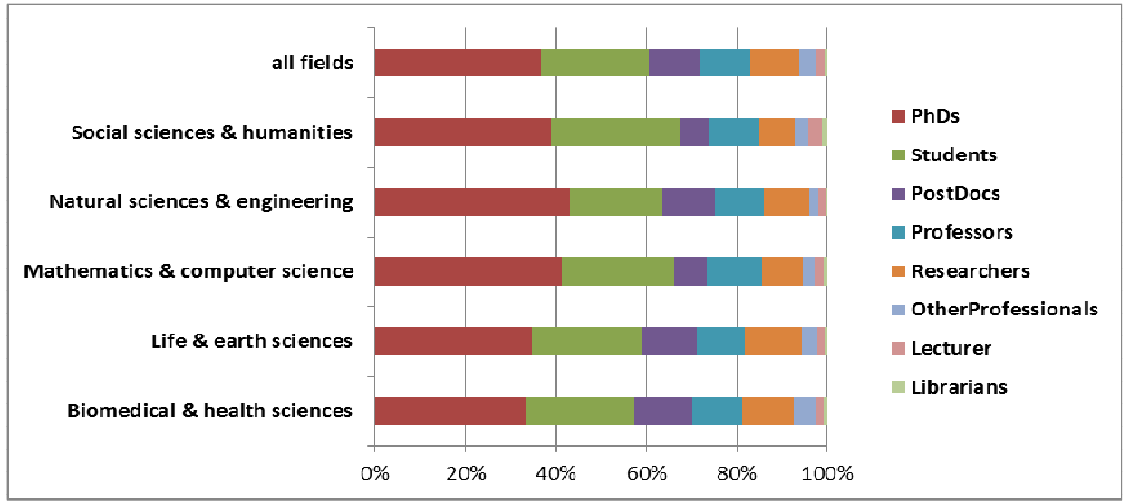

**Figure 1. Distribution of Mendeley readerships by the different types of users across LR fields**

*Relationship of Mendeley readerships with bibliometric indicators*
Spearman correlation analysis among readerships and bibliometric indicators and by the different types of users and across LR Fields has been calculated. The focus here is to explore the extent to which the readerships for the publications saved by the different users in Mendeley are related to their citations and journal indicators. Overall correlation scores among total readerships and bibliometrics indicators are positive and moderate ranging from p=.41 to p=.52 (Table 2).

**Table 2. Spearman Correlation analysis of bibliometrics and altmetrics variables**

| n=977,067 | CS | NCS | JCS | NJCS | RS |
|---|---|---|---|---|---|
| CS | 1 | .93 | .57 | .43 | .52 |
| NCS | | 1 | .40 | .46 | .50 |
| JCS | | | 1 | .75 | .44 |
| NJCS | | | | 1 | .41 |
| RS | | | | | 1 |

Citation Score (CS); Normalized Citation Score (NCS); Journal Citation Score (JCS); Normalized Journal Citation Score (NJCS); Readership Score (RS)

Regarding the different types of users, citations have a higher correlation with PhD followed by Students, PostDocs, Researchers, Professors and Other Professionals; however, Librarians and Lecturers exhibit the lowest correlations with citations. These different patterns in terms of correlations among the different types of users might suggest that they have different readership patterns and potentially different readership interests. For example, readership scores for Students, PostDocs, Professors and Researchers correlate most with PhD readership as 'Scientific users', which may indicate their similar scholarly and research usage behaviour. On the other hand, scientific users correlate less with 'other professionals' and Librarians (i.e. suggesting a kind of 'Professional users') and Lecturers as the 'Educational users' (Zahedi, Costas & Wouters, 2013). The latter also correlate most among themselves which may suggest both their similar use of scientific outputs and usage for other purposes than citation such as for self-awareness, teaching and educational or practical and professional purposes (Table3).

**Table 3. Spearman Correlation analysis of citation and readerships variables by types of Mendeley users**

| n=977,067 | CS | PhDs | Students | Post Docs | Professors | Researchers | Other Professionals | Lecturers | Librarians |
|---|---|---|---|---|---|---|---|---|---|
| **CS** | 1 | .46 | .40 | .41 | .36 | .37 | .24 | .18 | .06 |
| **PhDs** | | 1 | .58 | .49 | .48 | .47 | .25 | .27 | .08 |
| **Students** | | | 1 | .41 | .44 | .44 | .31 | .29 | .12 |
| **PostDocs** | | | | 1 | .42 | .43 | .26 | .21 | .06 |
| **Professors** | | | | | 1 | .39 | .27 | .26 | .09 |
| **Researchers** | | | | | | 1 | .32 | .23 | .11 |
| **Other Professionals** | | | | | | | 1 | .20 | .12 |
| **Lecturers** | | | | | | | | 1 | .09 |
| **Librarians** | | | | | | | | | 1 |

In terms of LR fields, the correlation of citations and readerships is the highest for Social sciences and humanities (p=.614) followed by Natural sciecnes and engineering (p=.597), Life and earth sciences (p=.578), Biomedical and health sciences (p=.553) and the least for Mathematics and computer sciences (p=.457). Regarding the readership by user types and across fields, for most users the highest correlations are in Social sciences and humanities. The lowest correlation with citations is in the field of Mathematics and computer sciences for PhD, Students, PostDocs, Professors and Researchers while for Other Professionals, Lecturers and Librarians the field Natural sciecnes and engineering displays the lowest correlation with citations (Table 4). This may indicate a relatively stronger use of social media platforms such as Mendeley by scholars in Social science and humanities in their research process than other fields (Rowlands et al. 2011; Tenopir, Volentine & King, 2013).

**Table 4. Spearman Correlation analysis of citation and readership by types of Mendeley users across 5 LR Fields**

| LR Fields | CS and RS | PhD | Student | Post Doc | Professor | Researcher | Other Professional | Lecturer | Librarian |
|---|---|---|---|---|---|---|---|---|---|
| **Biomedical & health sciences** | .54 | .47 | .42 | .42 | .40 | .39 | .26 | .19 | .05 |
| **Natural sciences &engineering** | .56 | .51 | .43 | .39 | .35 | .33 | .17 | .18 | .04 |
| **Life & earth sciences** | .56 | .53 | .46 | .43 | .40 | .39 | .24 | .22 | .06 |
| **Mathematics & computer science** | .43 | .42 | .34 | .26 | .26 | .27 | .18 | .18 | .05 |
| **Social sciences & humanities** | .60 | .54 | .50 | .41 | .43 | .42 | .31 | .27 | .12 |

CS (Citation Score); RS (Readership Score)

*Analyzing the filtering capacity of highly cited publications by Mendeley readerships*

The focus here is to explore the potential use of Mendeley users for filtering highly cited publications compared to journal citation scores. For this purpose, the proportion of top 10% highly cited publications (PPtop 10%)[9] in the sample have been detected. The precision-recall analysis[10] has been performed for all publications in the sample and the 5 LR fields and the different Mendeley users have been explored. Figure 2 shows the general precision-recall analysis of total readership scores and Journal Citation Scores (JCS) for all the publications in the dataset. This figure shows that readerships perform better than JCS in identifying the PPtop 10% most cited publications. The figure indicates that for example a recall of 0.5 (50%) corresponds with a precision of 0.45 (45%) for readership and 0.25 (25%) for journal citation scores in identifying highly cited publications, that is, publications belonging to the top 10% of their field in terms of citations. This means that in order to select half of all highly cited publications we have an error rate of 55% when the selection is made based on readership and an error rate of 75% when the selection is made based on journal citation scores. Since readership outperforms journal citation scores at all levels of recall, we conclude that readership scores identify highly cited publications much better than JCS.

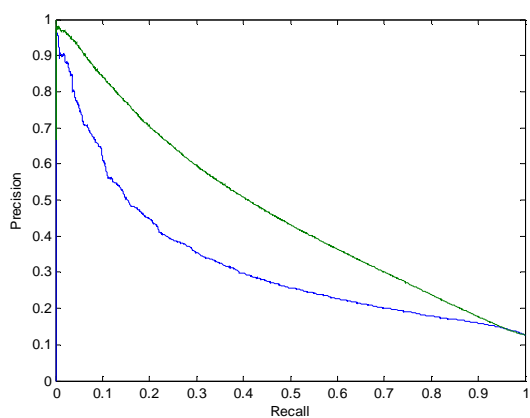

**Figure 2. General Precision-recall curves for JCS (blue line) and total readerships (green line) for identifying PPtop10% most highly cited publications**

*Precision-recall analysis of the different fields of science*
The results of the precision-recall analysis for all fields of science again show that readership outperforms JCS scores in filtering highly cited publications. This result supports the idea that Mendeley readership counts filter highly cited publications better than average citation impact of journals (JCS) for all LR fields within our sample. All the figures are similar resembling the general pattern in figure 2 except the figure for Mathematics & computer science which shows that from recall of 0.6 (60%), the two lines intersect each other and from that point onwards there is a small improvement of JCS over readership scores.

---

[9] PP(top 10%) (proportion of top 10% publications). Refers to the proportion of the publications that compared with other publications in the same field and in the same year, belong to the top 10% most frequently cited.

[10] following Waltman & Costas (2014), For a given selection of publications, "precision is defined as the number of highly cited publications in the selection divided by the total number of publications in the selection. Recall is defined as the number of highly cited publications in the selection divided by the total number of highly cited publications".

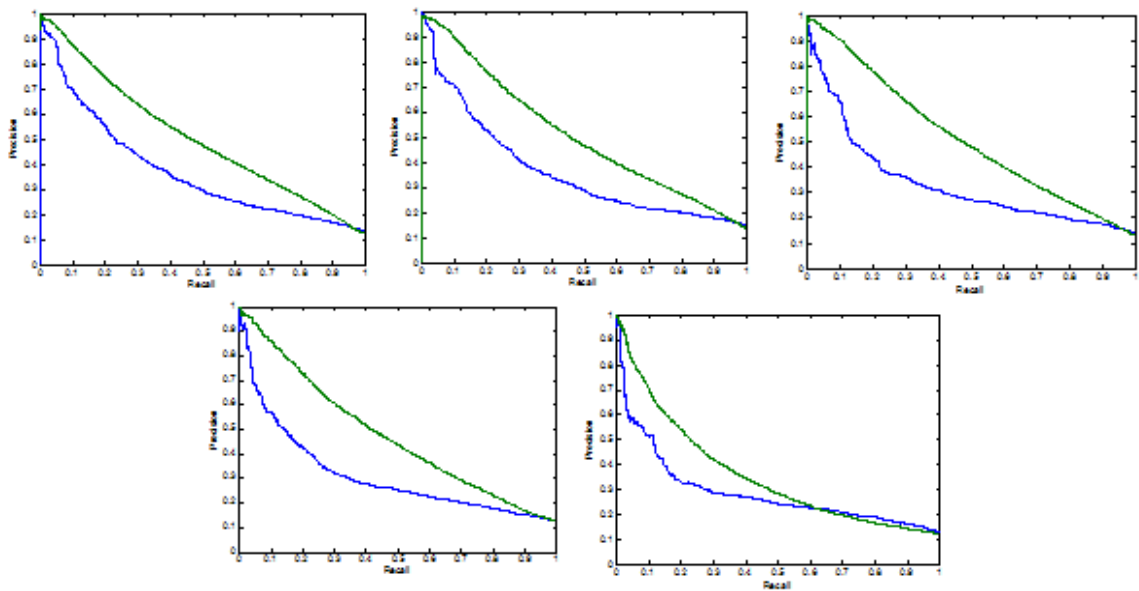

**From left to right: Biomedical & health sciences, Life & earth sciences, Natural Sciences & engineering, Social sciences & humanities, Mathematics & computer science**

**Figure 3. Precision-recall curves for JCS (blue line) and LR Fields (green line) for identifying PPtop10% most highly cited publications**

*Precision-recall analysis of different types of Mendeley users*
The same approach has been done based on the different Mendeley users. Figure 4 shows the results of the precision-recall analysis of readerships scores by the different types of users in Mendeley and Journal Citation Score (JCS). Again, readerships perform better than JCS for most types of users (PhDs, PostDocs, Professors, Researchers and Students vs Other Professionals, Librarians and Lecturers) in identifying the PPtop10% most highly cited publications within our dataset thus resembling the general pattern in Figure 2. The only exceptions are observed for Librarians, Lecturers and other Professionals where JCS overlaps or outperforms Mendeley readerships. This is in line with the result of the correlation analysis in which these Mendeley user types exhibit less correlations with citations than other types.
Also, regarding the figures for PostDocs, Professors, Researchers and Students, from recall point of 0.8 onwards two lines intersect each other and there is a slight improvement of JCS over readerships in the highest level of recall. However, in general, considering readership scores by most types of Mendeley users can help to detect highly cited publications.

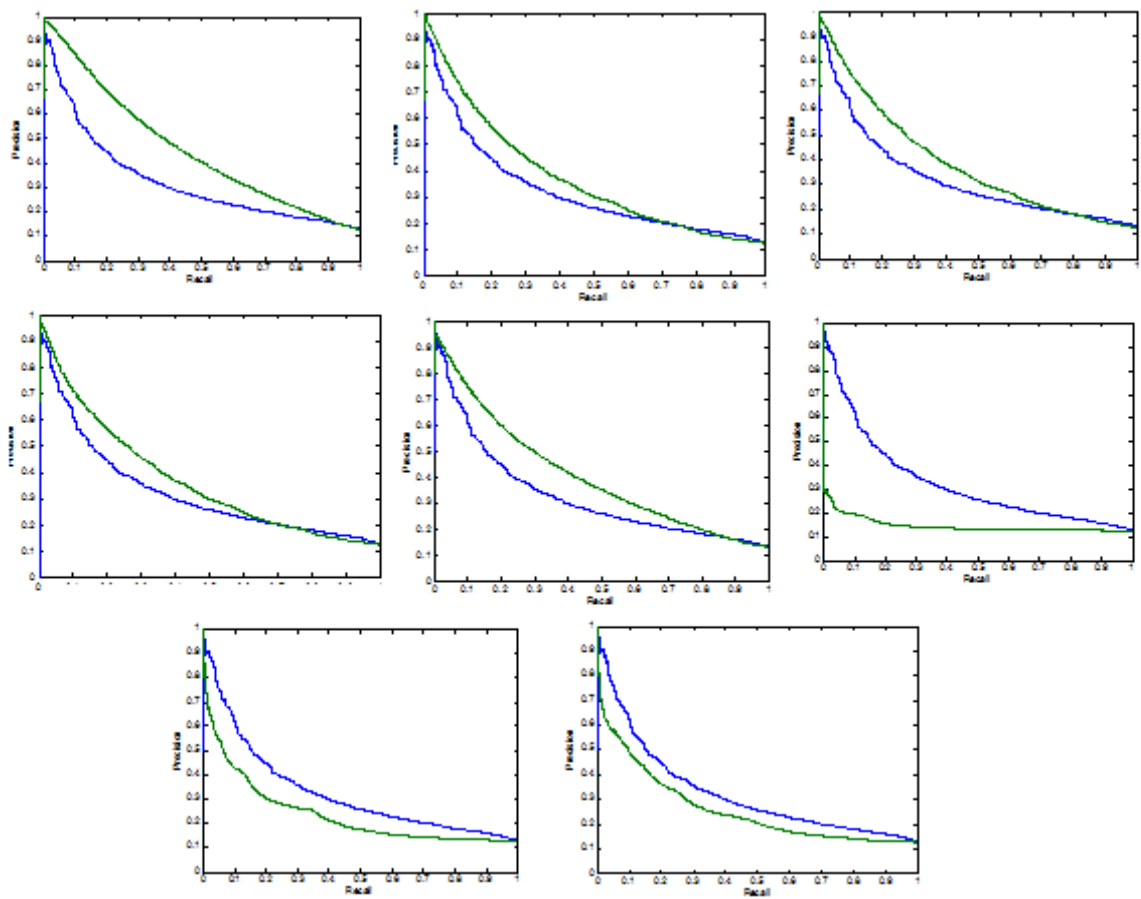

**From left to right: PhDs, PostDocs, Professors, Researchers, Students, Librarians, Lecturers and Other professionals**
**Figure 4. Precision-recall curves for JCS (blue line) and type of users readerships (green line) for identifying PPtop10% most highly cited publications**

**Main results and discussion**
Mendeley is a major multidisciplinary source of readership counts for scholarly publications (Zahedi, Costas & Wouters, 2014) and also it is one of the most promising tools for 'altmetrics' research (Li, Thelwall & Giustini, 2012; Wouters & Costas, 2012). The statistics about the 'Academic Status' of Mendeley users is a valuable source of information to learn more about the academic and non-academic positions of readers of scientific outputs, thus opening the possibility of studying the different types of impact that these different users may entail. Although Mendeley is now reporting the full data per publication, yet more clarity on how Mendeley users are defined is very important, as well as on how the typologies are chosen and updated by the users. For example, the relatively strong correlation between PhDs and Students could suggest that (some) students that become PhD do not update their profiles and therefore they 'read' like PhD students but without updating their 'Academic status' in Mendeley.

The current study has analysed and compared the readership and citation impact of the scholarly publications saved in Mendeley in terms of their types of users and across different LR fields, particularly focusing on the filtering capacity of readership and journal citation impact indicators in identifying highly cited publications. The findings showed that in terms of readership density across the 5 major LR fields, on average, all fields show higher MRS scores than MCS values. This suggests a faster reception of Mendeley readerships as compared to citations and encourages the need to study the temporality and pace of readership counts. Regarding the types of users, the most common types of users in Mendeley are PhDs and Students, for all LR fields. Correlation analysis shows relatively positive and moderate correlations among the different types of users and citations. The different correlations across users might support the idea that different users could be reading different publications, and thus justifying the use of 'Academic Status' to identify different reading behaviour and typologies of impact. For example, the higher correlations of scientific users with citations, supports their similar reading and citation behaviour vs. other more educational, teaching or professional patterns with lower correlations with citations. This may also be relevant in the analysis of the use of scientific publications in teaching or professional activities. Different correlation observed between particular research fields and types of users, reflecting the particular usage patterns of certain user as well as the general uptake of Mendeley in these fields.

Our results also suggest that readership counts really improve the filtering capacity of highly cited publications over JCS. This is one of the most promising results of this paper, showing the relevance of Mendeley readerships as a relevant filtering tool, something that has not been observed in the previous studies and for other altmetric sources (cf. Costas et al, 2014; Waltman & Costas, 2014). However, it should be taken into account that there are many scholars who don't use Mendeley or any other reference management tools in their scholarly process, so the act of using this type of tools may change in the future. Hence, the use of Mendeley readerships for evaluative purposes still needs careful consideration of its limitations and potential negative effects on the behaviour of individual scholars.

**Acknowledgments**
The authors are grateful to Erik van Wijk from CWTS for his support on the data collection of Mendeley information for this study. Also, special thanks to Ludo Waltman from CWTS for his fruitful comments on this paper. This work is partially supported by Iranian Ministry of Science, Research, and Technology (MSRT 89100156).